\documentclass[12pt]{spieman}  
\usepackage{amsmath,amsfonts,amssymb}
\usepackage{bm}
\usepackage{graphicx}
\usepackage{setspace}
\usepackage{tocloft}
\usepackage{color}
\usepackage{cases}
\usepackage{comment}

\title{Diffraction casting}
\author[a]{Ryosuke Mashiko}
\author[a,b]{Makoto Naruse}
\author[a,*]{Ryoichi Horisaki}

\affil[a]{The University of Tokyo, Graduate School of Information Science and Technology, Department of Information Physics and Computing, 7-3-1 Hongo, Bunkyo-ku, Tokyo, Japan, 113-8656}
\affil[b]{Deceased.}

\cftpagenumbersoff{figure}
\cftpagenumbersoff{table} 
\begin{document} 
\maketitle

\begin{abstract}
Optical computing is considered a promising solution for the growing demand for parallel computing in various cutting-edge fields, requiring high integration and high speed computational capacity.
In this paper, we propose a novel optical computation architecture called diffraction casting~(DC) for flexible and scalable parallel logic operations.
In DC, a diffractive neural network~(DNN) is designed for single instruction, multiple data~(SIMD) operations.
This approach allows for the alteration of logic operations simply by changing the illumination patterns.
Furthermore, it eliminates the need for encoding and decoding the input and output, respectively, by introducing a buffer around the input area, facilitating end-to-end all-optical computing.
We numerically demonstrate DC by performing all 16 logic operations on two arbitrary 256~bits parallel binary inputs. 
Additionally, we showcase several distinctive attributes inherent in DC, such as the benefit of cohesively designing the diffractive elements for SIMD logic operations, assuring high scalability and integration capability.
Our study offers a novel design architecture for optical computers and paves the way for a next-generation optical computing paradigm.

\end{abstract}


{\noindent \footnotesize\textbf{*}Ryoichi Horisaki,  \linkable{horisaki@g.ecc.u-tokyo.ac.jp} }

\begin{spacing}{2}   

\section{Introduction}
\label{sec:intro}  

Optical computing is a longstanding and captivating topic in the fields of optics and photonics.
It is considered a potential post-Moore computing technology~\cite{moore1998cramming}, offering distinct advantages including high bandwidth, rapid processing speed, low power consumption, and parallelism~\cite{caulfield2010future,kitayama2019novel}.
Around the 1980s, optical computing was actively explored, with developments in technologies such as optical vector matrix multipliers~\cite{goodman1978fully,bocker1974matrix,athale1982optical,gruber2000planar} and optical associative memories~\cite{owechko1987holographic,paek1987optical,ishikawa1989optical}.
Among these, shadow casting~(SC) emerged as a prominent optical computing technology of that era~\cite{tanida1983optical,ichioka1984optical,tanida1986opals,brenner1986digital}.
SC facilitated single instruction, multiple data~(SIMD) for logical operations through optical and spatially parallel computing.
The SC scheme relied on shadowgrams, which optically generated a single output image through massively parallel logic operations from two binary input images.
The versatility of SIMD logic operations was attained by altering the illumination pattern of the shadowgrams.
Another key aspect involved the computational encoding and decoding of input and output images, respectively, designed to balance light intensities between the zeros and ones in the binary images.
This computational process was an obstacle in achieving end-to-end optical computing.
Despite the anticipated benefits in speed and energy efficiency, these optical computing technologies in the 1980s stagnated due to limitations in hardware~(fabrication) and software~(design) for optical components at that time.
As a result, they lagged behind the major progress in electronic computing.

Over the past few decades, significant advancements in microfabrication, mathematical optimization, and computational power have dramatically transformed the field of optical computing from what it was in the 1980s.
Several pioneering optical computing techniques have been studied, including waveguide-based photonic circuits and diffractive neural networks~(DNNs).
Waveguide-based photonic circuits, which integrate waveguide interferometers, have high compatibility with currently existing electrical computers and circuits.
They have led to a wide range of applications, such as vector-matrix operations~\cite{miller2013self,shen2017deep,harris2018linear}, 
logical operations~\cite{zhang2005all,zhang2007optical,wu2008new,lee2008conceptual,dong2009proposal,fu2013silicon,sankar2020performance}, 
and integrated reconfigurable circuits~\cite{xie2017programmable,bogaerts2020programmable,ying2020electronic}.

DNNs consist of cascaded diffractive optical elements~(DOEs), which emulate neural network connections as light waves pass through the DOEs. 
This configuration utilizes the spatial parallelism of light and realizes fast and energy-efficient computation.
Extensive and attractive applications based on DNNs have been proposed, including image classification~\cite{lin2018all,yan2019fourier,chen2021diffractive,zhu2022space}, image processing~\cite{luo2022computational,icsil2022super,igarashi2023incoherent}, linear transformations~\cite{kulce2021all,li2022polarization,li2023massively} and logic operations~\cite{qian2020performing,wang2021orbital,zarei2022realization,luo2022cascadable,liu2023parallelized}.

Currently, the demand for computation in SIMD logic operations has intensified, particularly due to advancements in cutting-edge technologies such as image processing, machine learning, and blockchain~\cite{sonka2013image,dev2014bitcoin,reuther2019survey}.
Traditional computation with central processing units~(CPUs) is often inadequate to meet the computational needs of these fields.
Consequently, graphics processing units~(GPUs)~\cite{owens2008gpu}, tensor processing units~(TPUs)~\cite{jouppi2017datacenter}, field-programmable gate arrays~(FPGAs)~\cite{rose1993architecture,zhang2017machine}, and application-specific integrated circuits~(ASICs)~\cite{hari2015cryptocurrency,nurvitadhi2016accelerating} are employed as SIMD-specific devices. 
The trend towards high-speed, energy-efficient, and massively parallel computing aligns well with the advantages of optical computation.
As a result, there is a rapid increase in efforts to develop practical optical computing methods for SIMD logic operations.
Optical SIMD logic operations have been achieved using waveguide photonic circuits~\cite{ying2020electronic,bogaerts2020programmable}. 
However, a drawback of this approach is its limited scalability, which arises from the need for precise yet large-scale fabrication.
The use of DNNs holds potential as a solution to this issue, owing to the parallelism inherent in free-space propagation.
For instance, several types of DNN-based logic operations have been proposed to overcome this drawback~\cite{qian2020performing,wang2021orbital,zarei2022realization,luo2022cascadable,liu2023parallelized}, but the realization of SIMD logic operations using DNNs remains unachieved.
Moreover, these methods still require computational encoding and decoding of the input and output, respectively, posing a significant challenge toward end-to-end optical computing, similarly to the SC scheme.

In this study, considering the background mentioned above, we present a method termed diffraction casting~(DC) for conducting optical SIMD logic operations by incorporating the SC scheme and DNNs.
DC revives SC through the use of DNNs.
Therefore, DC shares the motivation of SC but exhibits several differences and advantages over SC.
Unlike SC, which is based on geometrical optics, DC is grounded in wave optics.
As a result, DC incorporates wave phenomena such as diffraction and interference through the use of DOE cascades in DNNs, and is anticipated to offer greater integration capability compared to SC.
Another advantage of DC is its elimination of the need for computational encoding and decoding of the input and output, respectively, which have been inherent bottlenecks in the SC scheme and previous DNN-based logic operations referred to above.
This is enabled by introducing a buffer area around the input pair.
In the rest of the paper, we will elaborate on the architectural design of DC, including the forward model, the optimization process, and provide numerical demonstrations.

\section{Materials and Methods}
\label{sec:Methods}

\subsection{Concept of Diffraction Casting}
\label{subsec:Methods_concept} 

\begin{figure}[]
\begin{center}
\includegraphics[scale=0.76]{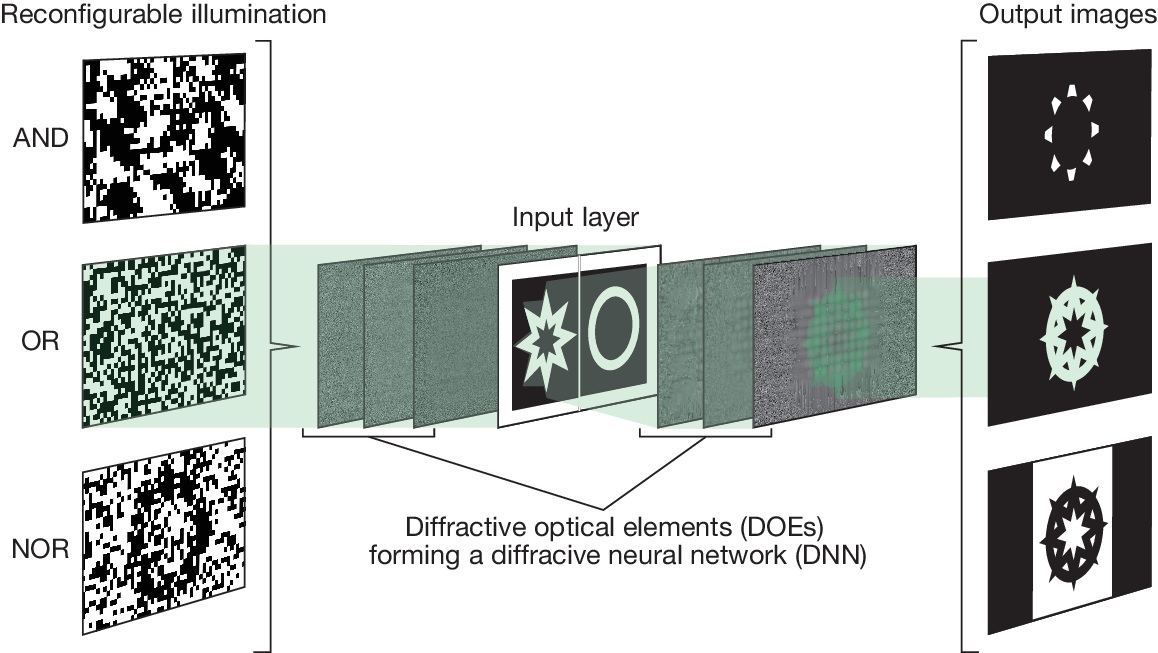} 
\end{center}
\caption 
{ \label{fig:Concept_thesis}
Schematic diagram of diffraction casting~(DC).
The selection of a logic operation is performed using reconfigurable illumination without any modification to the optical hardware.} 
\end{figure} 

DC is designed for 16 types of SIMD logic operations, processing two input binary images to produce one output binary image.
Figure 1 depicts the conceptual architecture of DC. 
DC consists of a reconfigurable illumination, DOEs, and an input layer.
The reconfigurable illumination enables the switch between logical operations and casts light on the DOE cascade forming a DNN.
In this paper, we focus on reconfigurable illumination with binary amplitude modulation and DOEs with phase modulation, assuming the use of commercially available optical components.
We place the two input images side by side on the input layer within the DOE cascade to achieve a simple optical setup.
The output of the logic operation appears as an intensity distribution at the end of the cascade and is captured with an image sensor.
The final result is binarized by assuming a one-bit image sensor or a computational process.
The reconfigurable illumination and DOEs are specifically trained to perform the 16 SIMD logic operations on any two binary images, as detailed in the subsequent subsection.
Once the training process is completed, DC enables massively parallel optical logic operations on arbitrary binary inputs just by selecting the illumination patterns, without necessitating any modifications to the DOEs.

\subsection{Optical Forward Model}
\label{subsec:Methods_forwardmodel} 
\begin{figure}[]
\begin{center}
\includegraphics[scale=0.76]{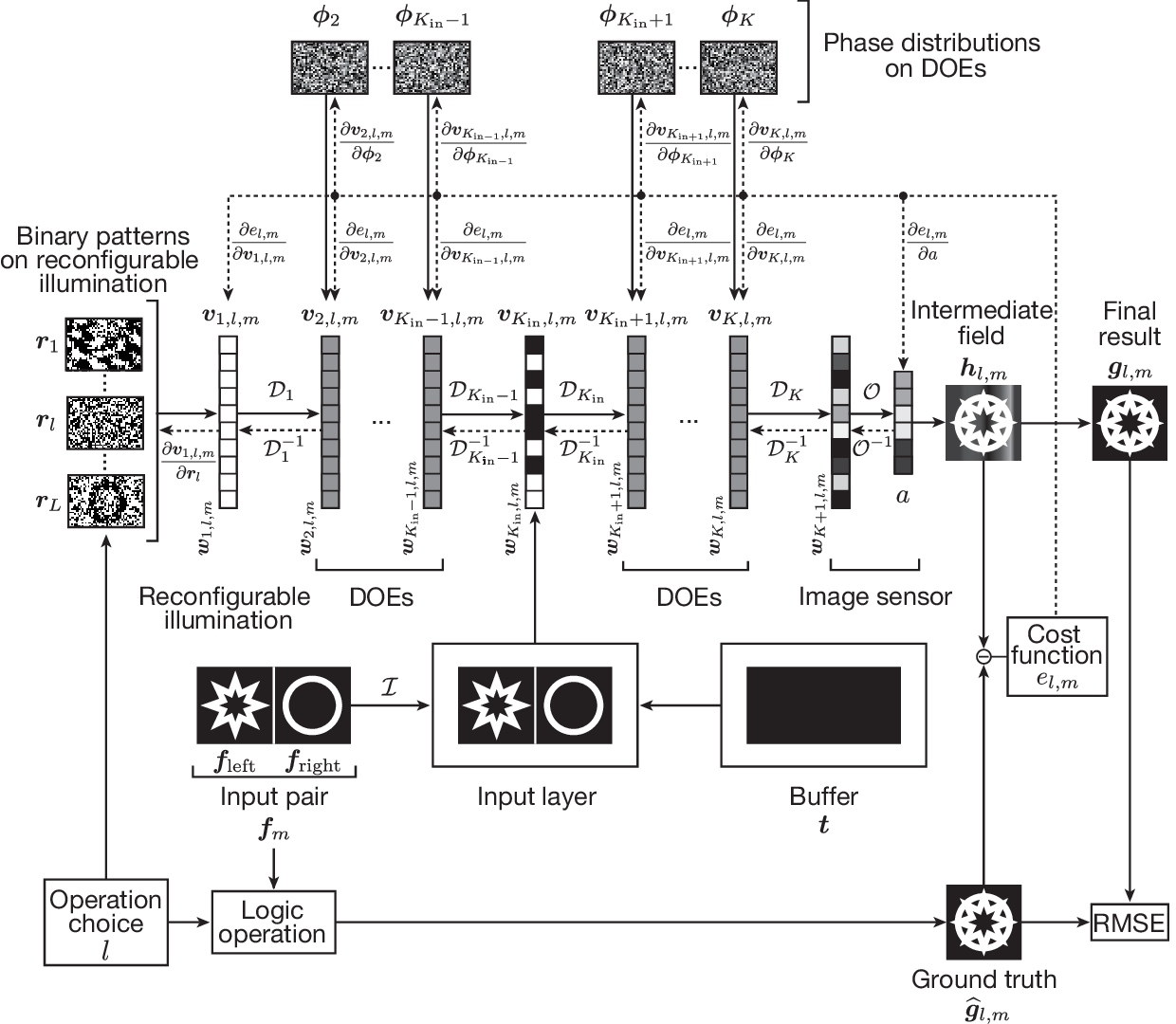} 
\end{center}
\caption 
{ \label{fig:Model_detailed}
Forward and backward processes of DC.
The reconfigurable illumination, the DOEs, and the scaling factor are optimized through the training process.}
\end{figure} 

Figure~\ref{fig:Model_detailed} illustrates the forward and backward processes of DC.
We consider a total $L$~types of SIMD logic operations on $N$~parallel bits.
These operations are conducted by the optical cascade composed of $K$~layers, including one illumination layer, one input layer, and $(K-2)$~DOE layers, with the layer index denoted as $k\in\{1,2, \ldots, K\}$.
The first layer of the optical cascade is the binary reconfigurable illumination~$\bm{r}_{l} \in \{0,1\}^{P_x \times P_y}$, where $l \in \{1, 2, \cdots, L\}$ is the index of the logic operations.
An input pair~$\bm{f} \in \{0,1\}^{N_x \times 2N_y}$, composed of side-by-side binary images, is located on the input layer, denoted as the $K_\mathrm{in}$-th layer in the cascade.

Here, $N_x$ and $N_y$ represent the pixel counts of the individual images within the input pair along the $x$- and $y$-directions, respectively, where $N_x \times N_y = N$.
The phase distributions for each DOE layer are denoted by $\bm{\phi}_k \in \mathbb{R}^{P_x \times P_y}$, and $P_x$ and $P_y$ indicate the pixel counts of the DOEs along the $x$- and $y$-axes, respectively.
The result of the logic operations is observed with the image sensor located downstream of the $K$-th layer in the optical cascade.

We describe below the forward process of DC.
The complex amplitude modulation~$\bm{v}_k \in \mathbb{C}^{P_x \times P_y}$, induced by the reconfigurable illumination, input pair, and the phase-only DOEs at the $k$-th layer in the cascade, is expressed as
\begin{numcases}
{\bm{v}_k=}
\bm{r}_{l}&for~$k= 1$,\label{eq_cascade_ilm}\\
\mathcal{I}[\bm{f}]+\bm{t}&for~$k=K_{\mathrm{in}}$,\label{eq_cascade_in}\\
\exp(j\bm{\phi}_k)&otherwise,\label{eq_cascade_doe}
\end{numcases}
where $j$ denotes the imaginary unit.
Here, $\mathcal{I}$ is an operator transforming the input pair into an amplitude image on the input layer, composed of following two steps.
The first step is up-sampling along the $x$- and $y$-directions with factors of $s_x\in\mathbb{N}$ and $s_y\in\mathbb{N}$, respectively.
The second step is zero-padding to enlarge the up-sampled input pair to the DOE size~($P_x\times P_y$~pixels).
$\bm{t}\in\{0,1\}^{P_x \times P_y}$ expresses a buffer surrounding the up-sampled input pair as illustrated in Fig.~\ref{fig:Model_detailed} and is defined as follows:
\begin{numcases}
{\bm{t}(u_x,u_y)=}
0&for~$\{u_x,u_y\}\in\Big(\frac{P_x-s_{x}N_{x}}{2},\frac{P_x+s_{x}N_{x}}{2}\Big]\times\Big(\frac{P_y-2s_{y}N_{y}}{2},\frac{P_y+2s_{y}N_{y}}{2}\Big]$,\label{eq_buffer_center}\\
1&otherwise,\label{eq_buffer}
\end{numcases}
where $u_x\in\mathbb{N}$ and $u_y\in\mathbb{N}$ are indices along the $x$- and $y$-directions.
This buffer is employed to compensate for light intensities transmitted or blocked on the input pair and enables the removal of computational encoding and decoding of the input and output, processes that are indispensably employed in previous optical logic operation methods, including the SC scheme.

The propagation process passing through the $k$-th layer in the cascade is written as
\begin{equation}
\bm{w}_{k+1} = \mathcal{D}_k[\bm{v}_k\bm{w}_k],
\label{eq_prop}
\end{equation}
where $\bm{w}_k \in \mathbb{C}^{P_x \times P_y}$ is a complex amplitude field just before the $k$-th layer.
$\mathcal{D}_k$ is a diffraction operator representing the propagation from the $k$-th layer to the $(k+1)$-th layer, calculated based on the angular spectrum method~\cite{goodman2005introduction}.
The initial field~$\bm{w}_1$ is specified as an all-one matrix, indicating a uniform field at the start.

The output intensity field of the optical cascade is observed with the image sensor as follows:
\begin{align}
\bm{h}&=a\mathcal{O}[|\bm{w}_{K+1}|^2],\label{eq_samp}\\
\bm{g}&=\mathcal{B}[\bm{h}].\label{eq_bin}
\end{align}
Here, $\bm{h}\in\mathbb{R}^{P_x\times P_y}$ represents the intermediate field, obtained through an operator~$\mathcal{O}$ that first crops the central $s_{x}N_{x}\times s_{y}N_{y}$ pixels from the output intensity field and then down-samples it to the original input image size of $N_x\times N_y$. 
This process includes scaling the intensity by a factor of $a\in\mathbb{R}_{>0}$, which corresponds to either amplifying or attenuating the signal. 
$\bm{g}\in\mathbb{R}^{N_x\times N_y}$ is the final result of the logic operation, with the binarization operator~$\mathcal{B}$ defined as
\begin{numcases}
{\mathcal{B}[z]=}
0&for~$z<0.5$,\label{eq_bin_0}\\
1&for~$z\geq 0.5$,\label{eq_bin_1}
\end{numcases}
where $z\in\mathbb{R}$ is an arbitrary variable. 
The binarization process, which converts analog signals to boolean ones, is implemented using either a one-bit image sensor or through computational means.

\subsection{Optimization Process}
\label{subsec:Methods_optimization}

To realize optical logic operations in parallel, the illumination~$\bm{r}_l$, the DOEs~$\bm{\phi}_k$, and the scaling factor~$a$ are optimized based on gradient descent in this study.
First, we describe the optimization process by assuming a single logic operation~($L=1$) and a single input pair for simplicity, where the illumination is defined as~$\bm{r}_{\scriptstyle{L=1}}$.
Then, we extend the optimization process to arbitrary numbers of logic operations and input pairs. 

\subsubsection{Derivatives for a single logic operation and a single input pair}
\label{subsubsubsubsec:optimization_formulation_unit}

We define a cost function~$e$ for a single logic operation and a single input pair as follows:
\begin{equation}
\label{eq_cost_unit}
e=\frac{1}{N}\sum_{\forall} \left|\bm{e}\right|^2,
\end{equation}
where $\displaystyle \sum_{\forall}$ represents the summation of all the elements of a tensor on its right side.
Here, the error~$\bm{e}$ is defined by
\begin{equation}
\label{eq_error_unit}
\bm{e}=\bm{h}-\widehat{\bm{g}},
\end{equation}
which represents the difference between the intermediate field~$\bm{h}$ and the ground truth of the operation result~$\widehat{\bm{g}}$.
This is to avoid the intermediate field's signals around the threshold values in $\mathcal{B}$, ensuring a robust binarization process.

To optimize $\bm{r}_{\scriptstyle{L=1}}$ and $\bm{\phi}_k$ based on gradient descent, the partial derivatives of $e$ with respect to these variables are expressed by employing the chain rule as follows:
\begin{align}
\frac{\partial e}{\partial \bm{r}_{\scriptstyle{L=1}}} &=\frac{\partial \bm{v}_1}{\partial \bm{r}_{\scriptstyle{L=1}}}\cdot\frac{\partial e}{\partial \bm{v}_1},\label{eq_illum_chain_unit}\\
\frac{\partial e}{\partial \bm{\phi}_k} &=\frac{\partial \bm{v}_k}{\partial \bm{\phi}_k}\cdot\frac{\partial e}{\partial \bm{v}_k}.\label{eq_doe_chain_unit}
\end{align}
The right sides of these three partial derivatives include the partial derivative of $e$ with respect to $\bm{v}_k$, calculated as:
\begin{equation}
\frac{\partial e}{\partial \bm{v}_k}
= \frac{4a}{N}\bm{w}^*_{k} \mathcal{D}^{-1}_{k}[\bm{v}^*_{k+1}\mathcal{D}^{-1}_{k+1}[\cdots[\bm{v}^*_{K}\mathcal{D}^{-1}_{K}[\bm{w}_{K+1}\mathcal{O}^{-1}[\bm{e}]]]\cdots]],
\label{eq_grad_v_unit}
\end{equation}
where $\mathcal{D}^{-1}_{k}$ and $\mathcal{O}^{-1}$ are operators representing the inverse processes of $\mathcal{D}_{k}$ and $\mathcal{O}$, respectively, and the superscript~$*$ denotes the complex conjugate.

The partial derivatives with respect to each optimized variable are finally written as follows:
The partial derivative with respect to $\bm{r}_{\scriptstyle{L=1}}$ is described as
\begin{equation}
\frac{\partial e}{\partial \bm{r}_{\scriptstyle{L=1}}} = \mathrm{Re}\left[\frac{\partial e}{\partial \bm{v}_1} \right],\label{eq_grad_illum_unit}
\end{equation}
where $\mathrm{Re}[\bullet]$ denotes the real part of a complex amplitude.
The partial derivative with respect to $\bm{\phi}_k$ is described as
\begin{equation}
\frac{\partial e}{\partial \bm{\phi}_k} = \mathrm{Re}\left[-j \bm{v}^*_k\frac{\partial e}{\partial \bm{v}_k}\right].
\label{eq_grad_doe_unit}
\end{equation}
The partial derivative with respect to $a$ is described as
\begin{align}
\frac{\partial e}{\partial a} 
= \frac{2}{aN}\sum_{\forall}\bm{h}\bm{e}.
\label{eq_grad_scale_unit}
\end{align}

\subsubsection{Derivatives for multiple logic operations and multiple input pairs}
\label{subsubsec:optimization_formulation}
Next, we extend the optimization process from a single logic operation and a single input pair as described above, to $L$~logic operations and $M$~input pairs. 
In this scenario, the cost function~$E$ is expressed as
\begin{equation}
E = {\frac{1}{LM}}\sum_{l,m}e_{l,m},
\label{eq_cost}
\end{equation}
where $e_{l,m}$ denotes the cost associated with the $l$-th logic operation and the $m$-th input pair, derived from Eq.~(\ref{eq_cost_unit}).
The partial derivatives of $E$ with respect to $\bm{r}_l$, $\bm{\phi}_k$, and $a$ are presented as summations of the partial derivatives of $e_{l,m}$ with respect to these variables, derived from Eqs.~(\ref{eq_grad_illum_unit})--(\ref{eq_grad_scale_unit}), respectively:
\begin{align}
\frac{\partial E}{\partial \bm{r}_{l}} &=\sum_{m}\frac{\partial e_{l,m}}{\partial \bm{r}_{l}},\label{eq_grad_illum}\\
\frac{\partial E}{\partial \bm{\phi}_k} &=\sum_{l,m}\frac{\partial e_{l,m}}{\partial \bm{\phi}_k},\label{eq_grad_doe}\\
\frac{\partial E}{\partial a} 
&=\sum_{l,m}\frac{\partial e_{l,m}}{\partial a}.\label{eq_grad_scale}
\end{align}

\subsubsection{Updating procedure}
\label{subsubsec:optimization_updating}

The variables~$\bm{r}_l$, $\bm{\phi}_k$, and $a$ are updated with the partial derivatives in Eqs.~(\ref{eq_grad_illum})--(\ref{eq_grad_scale}) based on the Adam optimizer~\cite{kingma2014adam}.
The updating processes for the DOEs~$\bm{\phi}_k$ and the scaling factor~$a$ are described as follows:
\begin{align}
\bm{\phi}_k&\leftarrow \bm{\phi}_k-\text{Adam}\left[\frac{\partial E}{\partial \bm{\phi}_k}\right],\label{eq_update_doe}\\ 
a&\leftarrow a-\text{Adam}\left[\frac{\partial E}{\partial a}\right],\label{eq_update_scale}
\end{align}
where $\text{Adam}\left[\cdot\right]$ represents an operator to calculate the updating step in the Adam optimizer with the derivatives.
To simplify the physical realization of the illumination~$\bm{r}_{l}$, we assume its binary implementations, such as digital micromirror devices, by introducing stochastic perturbations into the gradient descent process~\cite{courbariaux2015binaryconnect}.
The first step in the update process for the variables is as follows:
\begin{equation}
\tilde{\bm{r}}_{l}\leftarrow \mathcal{C}\left[\tilde{\bm{r}}_{l}-\text{Adam}\left[\frac{\partial E}{\partial \bm{r}_{l}}\right]\right],\label{eq_update_ilm_back}\\
\end{equation}
where $\tilde{\bm{r}}_{l}$ is an intermediate variable for the backward process in the optimization of $\bm{r}_l$.
Here, $\mathcal{C}$ is an operator for clipping the range of values as follows:
\begin{numcases}
{\mathcal{C}[z] =}
0 & for $z<0$,\label{eq_clip_min}\\
1 & for $z>1$,\label{eq_clip_max}\\
z & otherwise.\label{eq_clip_other}
\end{numcases}
Subsequently, $\bm{r}_{l}$ in the forward process is updated as follows:
\begin{equation}
\bm{r}_{l}=\mathcal{B}[\tilde{\bm{r}}_{l}+\bm{q}].
\label{eq_update_ilm_forward}
\end{equation}
where $\bm{q} \in \mathbb{R}^{P_x \times P_y}$ is a uniform distribution between $\pm 0.5$, introduced to avoid local minima in the binary optimization.
After the optimization process, $\bm{r}_{l}$ is finalized as follows:
\begin{equation}
\bm{r}_{l}=\mathcal{B}[\tilde{\bm{r}}_{l}].
\label{eq_update_ilm_final}
\end{equation}

\section{Numerical Demonstration}
\label{sec:num}
\subsection{Experimental Conditions}

\begin{table}[ht]
\caption{Logic operations defined on input pair.}
\label{tab:logic}
\begin{center}
\begin{tabular}{|c|c|c|c|} \hline
  Input pair&Operation&Logic&Boolean\\
   $\bm{f}_\text{left}$&index&operation& 0~0~1~1\\ 
   $\bm{f}_\text{right}$&$l$&&0~1~0~1\\ \hline
   Output& 1& 0& 0~0~0~0  \\ 
  $\widehat{\bm{g}}_{l}$&2& $\bm{f}_\text{left}\land \bm{f}_\text{right}$~(AND)& 0~0~0~1 \\ 
   &3 & $\bm{f}_\text{left}\land \overline{\bm{f}_\text{right}}$& 0~0~1~0\\
   &4 & $\bm{f}_\text{left}$& 0~0~1~1\\
   &5 & $\overline{\bm{f}_\text{left}}\land \bm{f}_\text{right}$ & 0~1~0~0\\
   &6 & $\bm{f}_\text{right}$& 0~1~0~1 \\
   &7 & $\bm{f}_\text{left}\oplus \bm{f}_\text{right}$~(XOR)& 0~1~1~0\\
   &8 & $\bm{f}_\text{left}\lor \bm{f}_\text{right}$~(OR)& 0~1~1~1\\ 
   &9& $1$&1~1~1~1\\
   &10 & $\overline{\bm{f}_\text{left}\land \bm{f}_\text{right}}$~(NAND)& 1~1~1~0\\
   &11& $\overline{\bm{f}_\text{left}}\lor \bm{f}_\text{right}$& 1~1~0~1 \\
   &12 & $\overline{\bm{f}_\text{left}}$& 1~1~0~0\\
   &13 & $\bm{f}_\text{left}\lor \overline{\bm{f}_\text{right}}$& 1~0~1~1\\
   &14 & $\overline{\bm{f}_\text{right}}$& 1~0~1~0\\
   &15 & $\overline{\bm{f}_\text{left}\oplus \bm{f}_\text{right}}$~(XNOR)& 1~0~0~1\\
   &16 & $\overline{\bm{f}_\text{left}\lor \bm{f}_\text{right}}$~(NOR)&  1~0~0~0\\
\hline
\end{tabular}
\end{center}
\end{table}
We numerically demonstrated DC with all sixteen logic operations~$(L=16)$ for the boolean input pair of $\bm{f}_\text{left}$ and $\bm{f}_\text{right}$, as shown in Tab.~\ref{tab:logic}.
In this numerical demonstration, the wavelength of the coherent light for the reconfigurable illumination~$\lambda$ was defined as 0.532~\textmu m.
The optical cascade comprised eleven layers~$(K=11)$, incorporating nine DOEs, with the input layer positioned as the sixth layer~$(K_{\mathrm{in}}=6)$.
The intervals between the layers were equally set to $3\times10^4\lambda~(\approx 1.60\times 10^4~\text{\textmu m})$.
For the illumination pattern~$\bm{r}_l$, the DOEs~$\bm{\phi}_k$, and the input layer, the pixel pitch was $16\lambda~(\approx8.51~\text{\textmu m})$, and the pixel count was $160~(=P_x)$ along the $x$-axis and $288~(=P_y)$ along the $y$-axis, respectively.
For the input pair~$\bm{f}_m$, the pixel count of the individual image in the pair was $16~(=N_x)$ along the $x$-axis and $16~(=N_y)$ along the $y$-axis, where the parallel bits~$N$ became 256, and the up-sampling factors~$s_x$ and $s_y$ were both 8, respectively.
The width of the region with ones on the buffer~$\bm{t}$, as defined in Eq.~(\ref{eq_buffer}), was set to 16~pixels.
To prevent the circulant effect on the diffraction calculation, the complex amplitude fields were zero-padded with a width of 64~pixels during the layer-by-layer propagation processes.

For the optimization process, input pairs were generated with values initially selected from uniform random distributions between 0 and 1, and then binarized using randomly selected thresholds also between 0 and 1.
The batch size~$M$ was set to 16 for training.
The number of the iterations was $5000$.
The learning rates for the Adam optimizer, as used in Eqs.~(\ref{eq_update_doe})--(\ref{eq_update_ilm_back}), for $\bm{r}_l$, $\bm{\phi}_k$, and $a$ were set to $3\times10^{-2}$, $1\times10^{-2}$, and $3\times10^{-3}$, respectively.
These variables were initially set to uniform random distributions for $\bm{r}_l$ and $\bm{\phi}_k$, and 10 for $a$.
The computation performance of DC was evaluated by the root mean squared errors~(RMSEs) between the final result~$\bm{g}_{l,m}$ and the ground truth~$\widehat{\bm{g}}_{l,m}$ for 256 randomly generated test input pairs.

\subsection{Result}
\label{subsec:Result} 

\begin{figure}[]
\begin{center}
\includegraphics[scale=0.76]{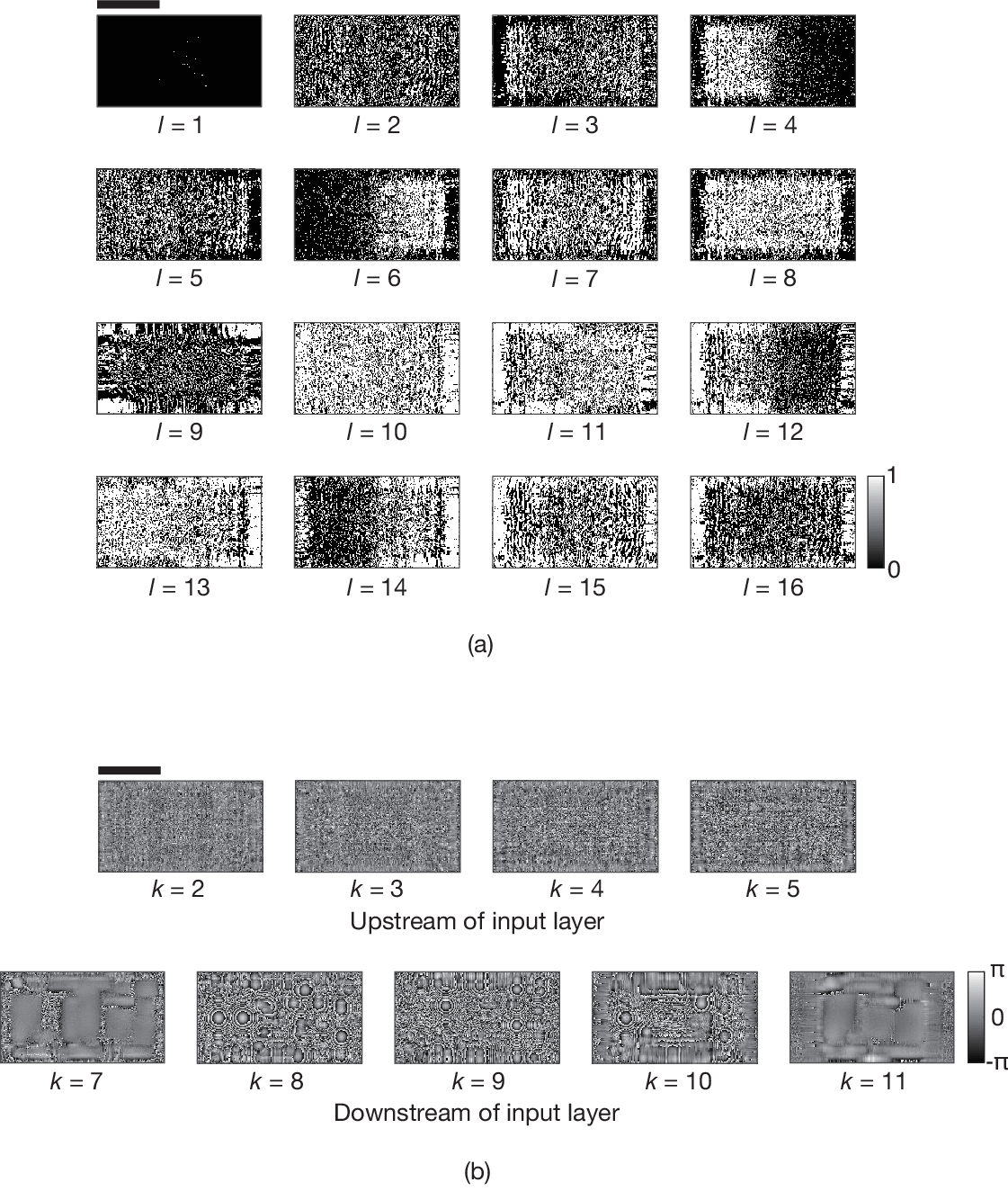} 
\end{center}
\caption 
{\label{fig:Result_optimized_variables} Optimization results.
(a)~Binary amplitude patterns on the reconfigurable illumination and (b)~phase distributions on the DOEs. Scale bar is 1~mm.}
\end{figure} 

\begin{figure}[]
\begin{center}
\includegraphics[scale=0.76]{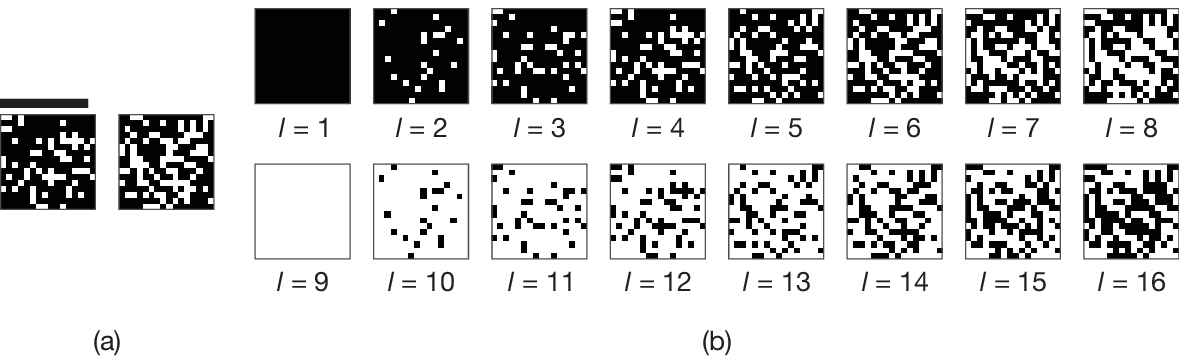} 
\end{center}

\caption{ \label{fig:Result_16operations}
An example of the DC process with the optimized illumination and the DOEs shown in Fig.~\ref{fig:Result_optimized_variables}.
(a)~Input pair and (b)~operation outputs.
Scale bar is 1~mm, indicating the physical scale after the up-sampling process.}
\end{figure} 

The optimization results for the illumination~$\bm{r}_l$ and the DOEs~$\bm{\phi}_k$ are presented in Figs.~\ref{fig:Result_optimized_variables}(a) and \ref{fig:Result_optimized_variables}(b), respectively.
The scaling factor~$a$ was optimized to 23.8.
Following this optimization, DC was numerically performed using an input pair with 256~parallel bits, which were not included in the training dataset, as shown in Fig.~\ref{fig:Result_16operations}(a).
Here, the sixteen SIMD logic operations were succeeded, as shown in Fig.~\ref{fig:Result_16operations}(b).
Moreover, the root mean squared errors~(RMSEs) between the final result~$\bm{g}_{l,m}$ and the ground truth~$\widehat{\bm{g}}_{l,m}$ for 256 randomly generated test input pairs were found to be 0.
These outcomes underscore the promising potential for DC.
More detailed discussions are provided in the subsequent section.

\section{Comprehensive Analysis}
\label{sec:CA} 
We conducted a numerical analysis of the performance of DC under various optical conditions.
This analysis evaluated the number of DOEs, the physical volume of DC, the position of the input layer, the energy efficiency, the buffer width, and the advantage of multiplexing operations.
Throughout this analysis, the experimental conditions were consistent with those described in Sec.~\ref{sec:num}, except where otherwise noted.

\subsection{Number of the DOEs}
\label{subsec:CA_doe} 

\begin{figure}[]
\begin{center}
\includegraphics[scale=0.76]{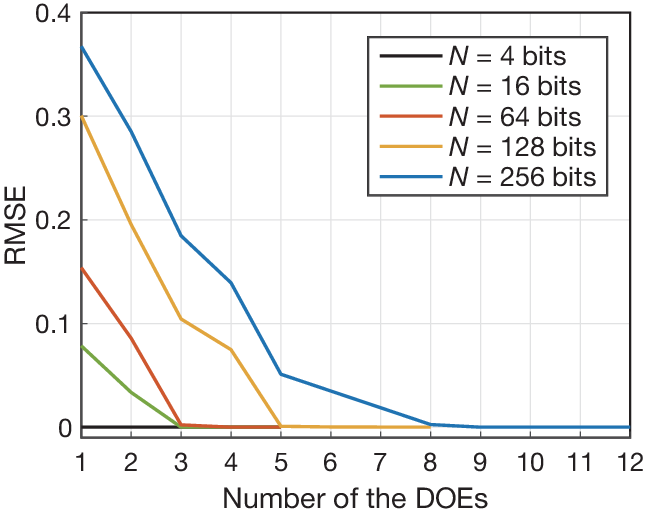} 
\end{center}
\caption 
{ \label{fig:CA_doe}
Computational errors associated with varying numbers of the DOEs.
}
\end{figure} 

The computational performance of DC, with the number of the DOEs set to $K-2$, was evaluated using the RMSEs, as illustrated in Fig.~\ref{fig:CA_doe}.
In this evaluation, the input parallel bits~$N$ were varied as 4, 16, 64, 128, and 256. 
Correspondingly, the up-sampling factors along $x$- and $y$-axis were adjusted to (64, 64), (32, 32), (16, 16), (8, 16), and (8, 8)~$(=(s_x, s_y))$, respectively, aiming to maintain consistent pixel counts~128~$(=s_{x}N_{x}, s_{y}N_{y})$ on the input layer after up-sampling.
The layer index of the input layer~$K_\mathrm{in}$ was set to $\lfloor (K-2)/2 \rfloor+2$.

As illustrated in Fig.~\ref{fig:CA_doe}, the calculation error decreased with an increase in the number of DOEs.
Additionally, the necessary number of DOEs for achieving error-free calculation increased with the input parallel bits~$N$, but at a rate less than proportional to $N$.
This rate of increase was smaller than predicted in previous works~\cite{kulce2021all,zhu2022space}, indicating an advantage of DC in terms of scalability and integration capability through the use of spatially parallelized optical processes for logic operations.

\subsection{Physical Volume of DC}
\label{subsec:CA_volume}
\begin{figure}[]
\begin{center}
\includegraphics[scale=0.76]{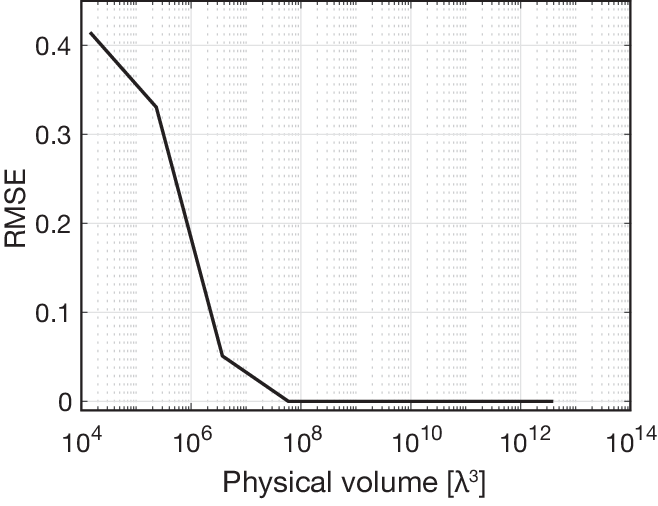} 
\end{center}
\caption 
{ \label{fig:CA_volume}
Computational errors associated with varying the physical volume of DC.
} 
\end{figure} 

The physical volume of the optical cascade in Sec.~\ref{subsec:Result} was calculated as $3.89\times10^{12}\lambda^3~(\approx 5.86\times10^{11}~\text{{\textmu m}}^3)$, where the pixel pitch on the DOEs was $16\lambda$ and the intervals between the layers in the optical cascade were $3\times 10^4\lambda$, respectively.
We investigated the computational error with respect to the reduction of the physical volume by scaling down both the pixel pitch and the interval with the same magnification ratio.
The result is shown in Fig.~\ref{fig:CA_volume}.
In this case, the minimal physical volume without computational error was $5.94\times10^7\lambda^3~(\approx 8.94\times10^6~\text{{\textmu m}}^3)$, where the pixel pitch was $\lambda~(=0.532~\text{\textmu m})$ and the interval was $1.17\times10^{2}\lambda~(\approx 6.23\times 10~\text{\textmu m})$, respectively.
This result indicated that the minimal physical volume of DC is constrained by the diffraction limit.

\subsection{Position of the Input Layer}
\label{subsec:CA_inputlayer} 
\begin{figure}[]
\begin{center}
\includegraphics[scale=0.76]{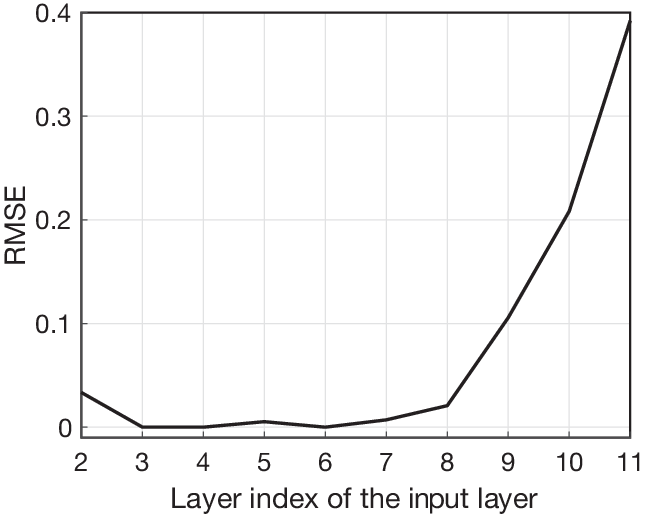} 
\end{center}
\caption 
{\label{fig:CA_Kin}
Computational errors under different positions of the input layer.}
\end{figure}

The computational error was calculated by varying the position of the input layer~$K_\mathrm{in}$ from 2 to 11 in the optical cascade with 11 layers, as depicted in Fig.~\ref{fig:CA_Kin}.
This result shows the importance of the DOEs downstream from the input layer in reducing computational error.
It suggests an advantage in positioning the input layer at an upper layer, excluding the top one.

\subsection{Energy Efficiency}
\label{subsec:CA_efficiency}

We evaluated the light energy efficiency of DC using the following definition:
\begin{equation}
\mathrm{Energy~efficiency}=\frac{\displaystyle\sum_{\forall}\left[\left.\mathcal{O}[|\bm{w}_{K+1}|^2]\right]\right|_{l=9,\bm{f}=1}}{P_xP_y}.
\label{eq_efficiency}
\end{equation}
Here, the denominator represents the total input energy to the optical cascade.
The numerator is the total energy on the output area of interest, calculated when the logic operation is configured to produce one output~($l=9$) and all elements of the input pair~$\bm{f}$ are set to one. 
The energy efficiency was assessed with respect to the scaling factor~$a$ and the width of the buffer~$\bm{t}$.

\subsubsection{Scaling factor}
\label{subsec:CA_scalingfactor}

\begin{figure}[]
\begin{center}
\includegraphics[scale=0.76]{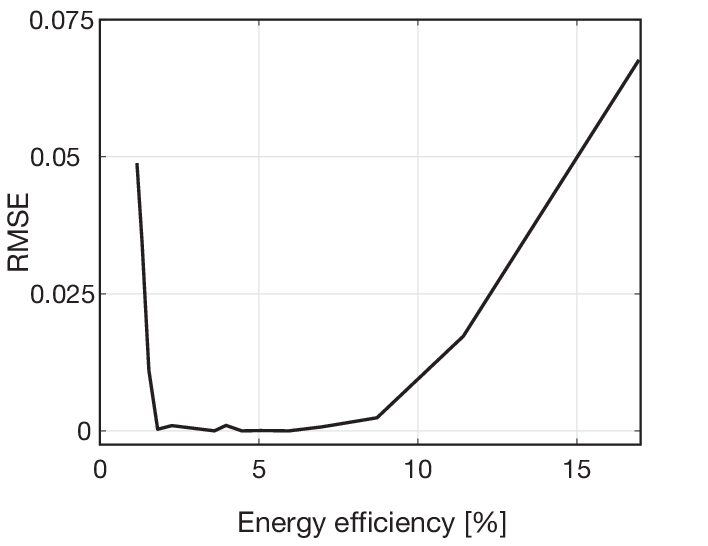} 
\end{center}
\caption 
{ \label{fig:CA_scalingfactor}
Relationship between computational errors and energy efficiencies when varying the scaling factor.
} 
\end{figure} 

The light energy efficiency is associated with the scaling factor~$a$, which amplifies or attenuates the signals captured by the image sensor before the binarization process.
A larger $a$ indicates lower energy efficiency and vice versa.
In the above demonstrations and analyses, $a$ was included in the optimized parameters, as shown in Eq.~(\ref{eq_update_scale}).
Here, $a$ was set to a specific value and was not updated during the optimization process.
Once the optimizations of the illumination pattern~$\bm{r}_l$ and the DOEs~$\bm{\phi}_k$ were completed, the energy efficiency in Eq.~(\ref{eq_efficiency}) and the computational error were calculated.
This process was repeated by changing $a$ from 2 to 32.
The relationship between the energy efficiency and the computational error is shown in Fig.~\ref{fig:CA_scalingfactor}.
The RMSEs for the energy efficiencies between 1.81\% and 8.71\% were less than $2.38\times 10^{-3}$.
Therefore, nearly error-free calculation was achieved within this range of energy efficiencies.

The primary sources of energy loss were amplitude modulation on the illumination plane, light leakage from the optical cascade, and the cropping of the limited square area by the image sensor at the end of the optical cascade.
The first issue can be solved by employing phase modulation on the illumination plane, although its modulation speed is lower than that of amplitude modulation on currently available spatial light modulators.
The second issue may be alleviated by reducing the intervals between layers.
The third issue can be addressed by increasing the sensor area, employing anisotropic sampling, or utilizing anamorphic imaging.
Another approach to improve the energy efficiency is to increase the width of the buffer, as indicated in the next section.

\subsubsection{Buffer width}
\label{subsubsec:CA_buffer}

\begin{figure}[]
\begin{center}
\includegraphics[scale=0.76]{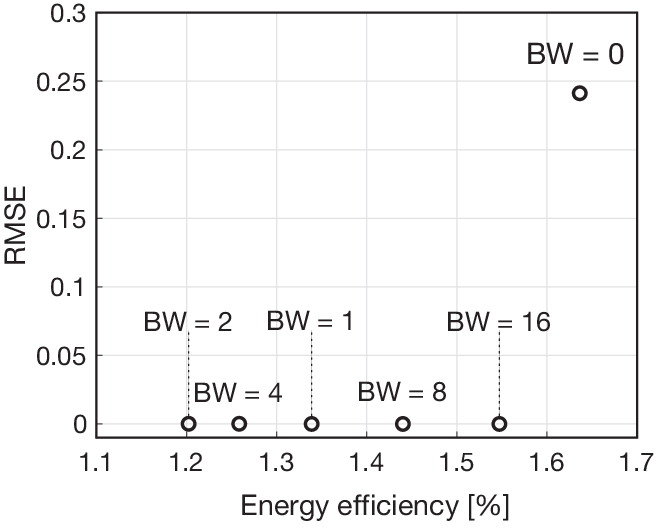} 
\end{center}
\caption 
{ \label{fig:CA_buffer}
Relationship between computational errors and energy efficiencies with varying buffer widths~(BW~[pixels]).}
\end{figure} 

The buffer~$\bm{t}$ was introduced into DC to compensate for the light intensities transmitted or blocked by the input pair.
It was expected to eliminate the computational encoding and decoding processes employed in previous methods for optical logic operations, including the SC scheme.
In the above demonstrations and analyses, the buffer width was set to 16~pixels.
The plots in Fig.~\ref{fig:CA_buffer} show the energy efficiencies and computational errors at different buffer widths, including zero width.
In this analysis, the scaling factor~$a$ was included in the optimization parameters.
This result supported the necessity of the buffer for error-free calculation.
Furthermore, a larger buffer width increased energy efficiency.

The RMSE for the logic operations at $1 \leq l \leq 8$ without the buffer was reduced from $4.24 \times 10^{-2}$ to 0 by adding a buffer of 1 pixel.
On the other hand, the RMSE for the logic operations at $9 \leq l \leq 16$ with no buffer was significantly improved from $4.40 \times 10^{-1}$ to 0 by adding a one-pixel buffer.
As shown in Tab.~\ref{tab:logic}, operations at $1 \leq l \leq 8$ do not include the operation with input of $\bm{f}_{\mathrm{left}} = 0, \bm{f}_{\mathrm{right}} = 0$ and output of $\widehat{\bm{g}}_l = 1$.
Conversely, operations at $9 \leq l \leq 16$ include such an operation.
This result also verified the role of the buffer---compensating for the balance between the light intensities of the input and output in the optical cascade.

\subsection{Multiplexing Advantage}
\label{subsec:CA_multiplex}

\begin{figure}[]
\begin{center}
\includegraphics[scale=0.76]{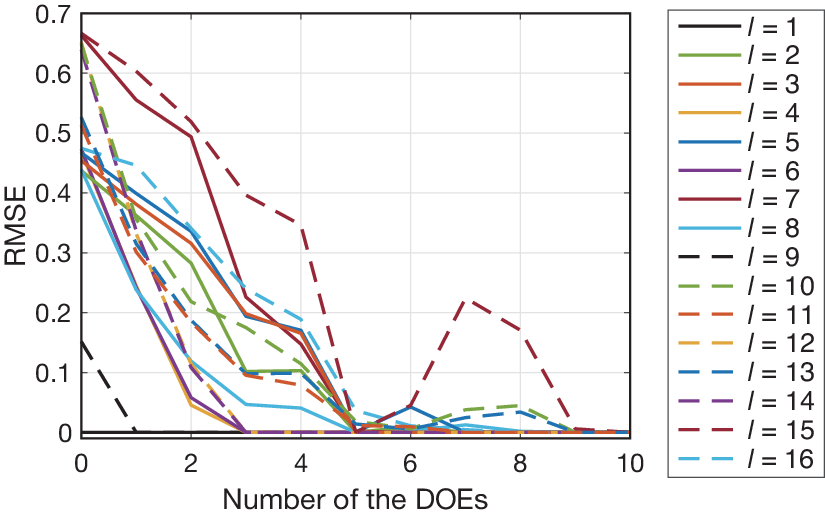} 
\end{center}
\caption 
{ \label{fig:CA_operationmultiplexity}
Computational errors associated with varying the number of DOEs for a single logic operation.}
\end{figure} 

The DOEs in DC multiplexed 16 logic operations, as shown in Tab.~\ref{tab:logic}.
We confirmed the computational errors under different numbers of DOEs, denoted as $K-2$, when the optical cascade was designed for a single logic operation, as illustrated in Fig.~\ref{fig:CA_operationmultiplexity}.
This highlights the benefits of multiplexing these operations.
The layer index of the input layer, $K_\mathrm{in}$, was set to $\lfloor (K-2)/2 \rfloor+2$.
When the number of DOEs was zero, only the illumination pattern was optimized for each logic operation.
In most cases, error-free calculations for single logic operations were achieved when the number of DOEs was larger than 6.
On the other hand, as shown in Fig.~\ref{fig:CA_doe}, the necessary number of DOEs for multiplexing 16 logic operations was 9, which is significantly less than $6\times 16$.
This result verified the multiplexing advantage and the integration capability of DC.

\section{Conclusion}
\label{sec:conclusion} 

We revived SC as DC by employing DNNs to achieve scalable and flexible optical SIMD operations.
The optical cascade of DC consisted of reconfigurable illumination, DOEs, and an input layer.
The illumination patterns and DOEs were designed to perform 16 logic operations on any binary input image pairs, and the output intensity of the optical cascade was binarized to produce the final results.
The advantages of DC included scalability, integration capability, and all-optical operation without the need for computational encoding and decoding, all of which were numerically demonstrated.

An issue with DC for practical applications is its low energy efficiency.
This may be addressed by adopting illumination with phase modulation and optimizing physical conditions, including the layer interval, buffer, and image sensor.
DC is extendable to a versatile mode of input and operation beyond the SIMD logic operations, owing to its flexible architecture and learning-based approach.
Furthermore, incorporating multiplexing in various optical quantities, such as time~\cite{zhang2024space,zhou2024spatiotemporal}, wavelength~\cite{duan2023optical}, polarization~\cite{li2022polarization,luo2022metasurface}, and orbital angular momentum~\cite{wang2021orbital} would enhance the computational capacity in DC.
Thus, our study on DC offers a novel design architecture for optical computers and optical accelerators, and paves the way for a next-generation optical computing paradigm.

\subsection*{Disclosures}
The authors declare no conflicts of interest.

\subsection*{Code and Data Availability}
Data may be obtained from the authors upon reasonable request.


\bibliography{report}   
\bibliographystyle{spiejour}   

\end{spacing}
\end{document}